\documentclass[12pt]{article}
\usepackage{graphicx,amsmath,amssymb,bm,multirow}
\DeclareMathOperator{\erf}{erf}
\def\be {\begin{equation}}
\def\ee {\end{equation}}
\def\ba {\begin{eqnarray}}
\def\ea {\end{eqnarray}}

\def\lb {\label}

\def\vf  {\varphi}

\def\bi {\begin{itemize}}
\def\ei {\end{itemize}}

\def\bea{\begin{eqnarray}}
\def\eea{\end{eqnarray}}
\begin{document}

\title{Scalar hairs and exact vortex solutions   in 3D AdS gravity}

\author{ Mariano Cadoni,\footnote{email: mariano.cadoni@ca.infn.it }\,
Paolo Pani\footnote{email: paolo.pani@ca.infn.it  }\,
and Matteo Serra \footnote{email:matteo.serra@ca.infn.it }\,
\\
{\it \small Dipartimento di Fisica, Universit\`a di Cagliari and
INFN, Sezione di Cagliari }\\}
\maketitle
\abstract{
We investigate three-dimensional (3D) Anti-de Sitter (AdS)
gravity coupled to a complex scalar 
field $\phi$ with self-interaction potential $V(|\phi|)$. 
We show that the mass of  static, rotationally symmetric,  AdS black hole 
with scalar hairs    is determined algebraically by 
the  scalar charges.
We recast  the field equations as a linear system of first order 
differential equations. Exact solutions, describing 3D AdS black holes 
with real spherical  scalar hairs and vortex-black hole solutions are derived in  
closed form  for the case of a scalar field saturating the
Breitenlohner-Freedman (BF) bound and for  a scalar 
field with asymptotic zero mass. The physical properties of these solutions 
are  discussed. In particular, we show that  the  vortex solution    
interpolates between two different AdS$_{3}$ vacua, 
 corresponding  respectively  to  a
$U(1)$-symmetry-preserving 
maximum and to a symmetry-breaking minimum of the potential $V$.}

\section{Introduction}
One of the most striking features of black hole solutions in any 
space-time dimension is their uniqueness. Typically, they are 
characterized by a  bunch of parameters, which are asymptotic  charges 
associated with global symmetries of the solution.
One question that has been debated since a long time is the 
uniqueness of  black hole solutions in the presence of scalar 
fields. The issue  is rather involved and a precise formulation 
of the  no-hair conjecture, i.e the 
absence  of  nontrivial scalar field configurations in 
a black hole background, has not yet been given  
\cite{Bekenstein:1972ny,Teitelboim:1972qx,
Bekenstein:1995un}. In this context the relevant question 
concerns  not only the presence of scalar hairs but also the possibility of 
having scalar charges independent from the black hole mass.

In recent years the investigation on black hole solutions with scalar 
hairs has  been mainly focused  on asymptotically AdS
solutions \cite{Hertog:2006rr}.  The obvious reason behind such interest is the 
anti-de Sitter/conformal field theory (AdS$_{d}$/CFT$_{d-1}$) correspondence.
The dynamics of scalar fields in the  AdS$_{d}$ background gives 
crucial  
information about the dynamics of the dual $(d-1)$-dimensional field 
theory. For instance, scalar hair configuration may interpolate between
two different AdS geometries (one asymptotic and the other 
near-horizon) \cite{Hotta:2008xt} similarly to what happens for charged AdS 
black holes \cite{Cadoni:2008mw,Cadoni:2008pr, Martinez:1999qi}. 

An other important example is given by the 
recently discovered holographic superconductors \cite{Hartnoll:2008vx,
Hartnoll:2008kx}.
It is exactly a nontrivial scalar hair black hole solution of 
four-dimensional (4D)
AdS gravity that breaks the $U(1)$ symmetry and is responsible for 
the superconducting phase transition in the dual three-dimensional (3D)
field theory.

No-hair theorems for asymptotically AdS$_{d}$ black hole have been  
discussed in  Ref. \cite{Hertog:2006rr} for the case $d>3$ . In that paper 
Hertog  has 
shown that AdS black hole solutions with spherical scalar hairs  can 
only  exist 
 if the positive 
energy theorem is violated. He was also able to construct numerical black hole 
solution with scalar hairs \cite{Hertog:2006rr}. This result has played an 
important role for the development of holographic superconductors. 
In fact, the  existence of 4D 
black holes dressed with a scalar hair is a necessary condition for 
having a phase transition in the dual theory.

In this paper we will consider the $d=3$ case. We will investigate the existence 
of black hole 
solutions with spherical scalar hairs in 3D AdS spacetime. Because  three 
spacetime dimensions may allow for topologically non trivial global 
vortex configuration for the scalar field, we consider the case of a 
complex scalar field. The interest for global vortex solutions is not 
only motivated by the search for nontrivial scalar hairs. The study 
of black holes formed by global vortex  is interesting by itself,  as 
a particular case of black hole solutions supported by solitons.
This kind of solutions may have some relevance in cosmology (as 
viable cosmic string candidates \cite{Kim:1997ba}), in the context of the 
AdS/CFT correspondence as nonperturbative 3D bulk solutions 
interpolating between different AdS vacua \cite{Hotta:2008xt} and also as 
possible gravitational duals of condensed matter systems \cite{Maity:2009zz}. 

As a first step we analyze the dynamics  of a complex  scalar field 
$\phi$ with an arbitrary potential $V(|\phi|)$  
in AdS$_{3}$ . We demonstrate a 
theorem stating that 
for  static, rotationally symmetric black hole solutions with scalar 
hairs   the  black hole mass is determined algebraically by 
the 
scalar charges. As a second step  we show that the field 
equations can be recast  in the form of a system of first order linear 
differential equations for the scalar potential $V$ and the spacetime 
metric. In principle, this allows us  to find  exact 
solutions of the field equations once the form of the scalar field is 
fixed. 
We use this method  to find exact black hole  solutions 
with spherical scalar hairs (both real and vortex-like)  in the case of  a scalar field 
saturating the BF bound \cite{Breitenlohner:1982bm}
and for  a scalar 
field with zero mass. 
These dressed black hole solutions are completely characterized by a scalar 
charge $c$ and a vortex winding number $n$, the black hole mass 
being determined by $c$ and $n$. 

In the case of a scalar field  saturating the BF bound, the potential 
$V(|\phi|)$ has a $W$, Higgs-like, form and a global 
$U(1)$ symmetry. We find that the black hole-vortex solutions 
interpolates between two AdS$_{3}$ geometries with different AdS 
lengths. Choosing appropriately the parameters of the potential these 
 AdS$_{3}$ geometries correspond respectively  to the 
$U(1)$-symmetry-preserving 
maximum and to the symmetry-breaking minimum of the Higgs potential.

The structure of the paper is as follows.
In Sect 2. we  discuss the dynamics of a complex scalar field in 3D 
AdS spacetime. In Sect. 3. we  discuss no-hair theorems  and
prove  the  theorem stating that the mass of  AdS$_{3}$ black holes 
with spherical scalar 
hairs is determined by $c$ and $n$.
In Sect. 4 we recast the field equations  in  linear form.
In Sect. 5 we derive and discuss  exact solutions of our system describing  
AdS$_{3}$ black holes endowed  with a real scalar hair saturating the 
BF bound. We also investigate 
their thermodynamics and show that their free energy is always 
higher then the free energy of the Banados-Teitelboim-Zanelli
(BTZ) black hole \cite{Banados:1992wn}.
In Sect. 6 we derive and discuss exact black hole-vortex solutions 
obtained for two different choices of the scalar field profile 
($|\phi|=c/r$ and $|\phi|=c/r^{2}¥$ ) 
and discuss the linear  stability of the BTZ  background.
Finally in Sect. 7 we present our concluding remarks.

\section{Dynamics of a complex scalar field in 3D AdS spacetime}

Let us consider 3D Einstein  gravity coupled to a
complex scalar field $\phi$:%
\begin{equation}
I=\frac{1}{2\pi }\int\, d^{3}x¥\sqrt{-g}\left[ R-V(|\phi|  )-%
\frac{1}{2}\partial _{\mu }\bar{\phi }\partial ^{\mu }\phi \right] ,
\label{4.1}
\end{equation}%
where $V(|\phi|)$ is the potential for the scalar and 
for convenience we set the 3D gravitational constant $G$ equal to $%
\frac{1}{8}$. The action (\ref{4.1}) is invariant under  global 
$U(1)$ transformations acting on the scalar field as $\phi \to e^{i 
\beta }\phi$.
We are only interested in solutions for the 3D metric that  are 
spherically symmetric and
asymptotically AdS. Therefore, we require that $V$ 
has a local 
extremum at $\phi=0$ (corresponding to $r=\infty$), such that 
\be\lb{cc}
V(0)=-\frac{2}{L^{2}},
\ee
where $L$ is the AdS length.
For a scalar field in the AdS$_{d}$ spacetimes stability 
of the asymptotic solution does not 
necessarily requires  $\phi=0$ to be a minimum.
Only the weaker BF bound, 
$m^{2}L^{2}\ge-(d-1)^{2}/{4}$, involving the squared mass $m^{2}¥=
V''(0)$ 
of the scalar (the prime denotes derivation with respect to $|\phi|$), 
must be satisfied. The scalar field must behave asymptotically as 
\footnote{When the BF bound is saturated the asymptotic 
behavior of the scalar field  may involve a logarithmic branch, 
whose back reaction  also modifies  the asymptotic behavior of the metric 
\cite{Henneaux:2004zi}. However, throughout this paper we will not consider 
this logarithmic branch.}
\be\lb{ae}
\varphi =\frac{ c_{-}}{r^{\lambda _{-}}}+\frac{c _{+}}{r^{\lambda _{+}}}%
+...,
\ee
where
$\lambda _{\pm }=\frac{1}{2}\left( d-1\pm 
\sqrt{(d-1)^{2}+4m^{2}L^{2}}\right).$

The field equations following from the action (\ref{4.1}) are,
\begin{eqnarray}\lb{fe}
R_{\mu \nu }-g_{\mu \nu }V(|\phi| )-\frac{1}{4}\partial _{\mu }%
\bar{\phi }\partial _{\nu }\phi-\frac{1}{4}\partial _{\mu }%
{\phi }\partial _{\nu }\bar\phi &=&0,\nonumber  \\
\frac{1}{2\sqrt{-g}}\partial _{\mu }(\sqrt{-g}\partial ^{\mu }\bar{\phi 
})-\frac{\partial V(|\phi| )}{\partial \phi } &=&0,\nonumber  \\
\frac{1}{2\sqrt{-g}}\partial _{\mu }(\sqrt{-g}\partial ^{\mu }\phi )-\frac{%
\partial V(|\phi| )}{\partial \bar{\phi }} &=&0.
\end{eqnarray}
 
We are interested in static, spherically symmetric solutions of  these equations, 
therefore we choose the following ansatz for the
metric:%
\begin{equation}\lb{ans}
ds^{2}=-e^{2f(r)}dt^{2}+e^{2h(r)}dr^{2}+r^{2}d\theta ^{2}.
\end{equation}%
The most general form of the complex scalar 
compatible with 
spherical symmetry of the metric is given in terms  of a real
modulus field $\vf(r)$ depending only on $r$ and  a phase depending linearly 
on $\theta$. This is a consequence of the global $U(1)$ symmetry  of the 
action (\ref{4.1}). For topologically nontrivial spacetimes the phase 
becomes physically relevant and the proportionality factor is quantized.
We will therefore use for $\phi$ the ansatz for  static global 
vortices:%
\be\lb{ans1}
\phi (r,\theta) = \vf(r) e^{in\theta }, \\
\ee
\bigskip where $n$ is the winding number  of the vortex. 

Using Eqs. (\ref{ans}) and (\ref{ans1}) in the field equations 
(\ref{fe}) we obtain:%
\begin{eqnarray}
V+\frac{1}{r}e^{-2h}{\dot f}+e^{-2h}{\dot f}^{ 2}-e^{-2h}{\dot
f}{\dot h }+e^{-2h}{\ddot f } &=& 0 \nonumber \\
-V+\frac{1}{r}e^{-2h}{\dot h }-e^{-2h}{\dot f}^{2}+e^{-2h}{\dot
f}{\dot h }-e^{-2h}{\ddot f}-\frac{1}{2}e^{-2h}{\dot \vf}^{2} &=& 
0 \nonumber \\
-V-\frac{1}{r}e^{-2h}{\dot f }+\frac{1}{r}e^{-2h}{\dot h }-\frac{1}{%
2}\frac{n^{2}}{r^{2}} \vf ^{2}&=& 0  \nonumber \\
- V^{\prime}¥+e^{-2h}\left( 
\frac{1}{r}+{\dot f}-{\dot h }\right) 
{\dot \vf }+e^{-2h}{ \ddot \vf  }-
\frac{n^{2}}{%
r^{2}} \vf  &=& 0,  \label{4.11}
\end{eqnarray}
where the  dot indicates the derivative with respect to the radial 
coordinate $r$.
Only three of the four Eqs. in (\ref{4.11}) are independent. For 
$\phi\neq const.$ the system can be drastically simplified by 
combining the first and second equation and 
multiplying the fourth by ${\dot \vf}$, 

\begin{eqnarray}
&&{\dot h }+{\dot f} =\frac{1}{2}r{\dot \vf} ^{ 2}  \nonumber  \\
&&{\dot h }-{\dot f} =r\left( V+\frac{1}{2}\frac{n^{2}}{r^{2}}\vf
^{2}\right) e^{2h} \nonumber  \\
&&e^{-2h}\left[\left( \frac{1}{r}+{\dot f }-{\dot h}\right)
{\dot\vf} ^{ 2}+\dot{\vf}\ddot{\vf}\right]  
-\frac{1}{2}\frac{n^{2}}{r^{2}}\dot {(\vf ^{2})}-{\dot V}=0  
\label{4.14}
\end{eqnarray}

\section{Scalar hairs}
In this section we will discuss no-hair theorems for the black 
solutions of 3D AdS gravity coupled with a scalar field.
The field equations (\ref{4.11}) admit as solution the 
BTZ black hole. This is obtained by 
setting $\vf=0$. Requiring  that $V^{\prime}(0)=0$ and using 
Eq. (\ref{cc}), one easily finds:
\be\lb{btz}
e^{2 f}=e^{-2h}= \frac{r^{2}}{L^{2}}-M,
\ee
where $M$ is the black hole mass.
One important question one can ask concerns  the existence of 3D AdS black 
holes with scalar hairs, i.e. solutions of the system (\ref{4.14}) 
with a non-constant scalar, $\vf\neq const.$
Furthermore, assuming such black holes with scalar hairs  to exist, 
one would also like to 
know whether  the scalar charges $c_{\pm}$ (see Eq. (\ref{ae}))
are  independent from the black hole mass 
$M$.

In the 4D case  it has been shown that the existence of AdS black 
holes with scalar hairs is crucially related with the violation of 
positive energy theorem \cite{Hertog:2006rr}. Generically,  this will 
be also  true 
for the 3D case. Therefore, we expect black holes with scalar hairs 
to exist
when the scalar mass squared $m^{2}$ becomes negative.
In the next sections we will derive in closed form, 
exact black hole solutions with scalar 
hairs in the case of $m^{2}$ saturating the BF bound and for $m^{2}=0$.
In this section we will be concerned with the second part of the 
 question above. We will show that the following  
theorem holds:

\textit{If we couple 3D AdS gravity with a complex scalar
field, the mass of static, rotationally symmetric solutions of the 
theory is determined algebraically 
by the scalar 
charges $c_{\pm}$ and by the winding number $n$. }
 
In order to demonstrate this theorem we first solve for $e^{2f}$ the system 
(\ref{4.14}).
One easily finds, 
\begin{equation}
e^{2f}= a_{0}e^{\int dr r {\dot \vf}^2}¥\left[ {\dot V}+
r\left( V+\frac{1}{2}\frac{n^{2}}{r^{2}}\vf
^{2}\right) {\dot\vf} ^{ 2}+\frac{1}{2}\frac{n^{2}}{r^{2}}\dot{(\vf
^{2})}\right] \left( \frac{1}{r}{\dot\vf} ^{2}+\ddot {\vf}
\dot{\vf}\right) ^{-1},  \label{542}
\end{equation}
where $a_{0}$ is an integration constant, which can be scaled away by 
a
rescaling of the time coordinate $t$. Because the $g_{tt}$ component 
of the metric 
is a function of  $V,\vf,n$ the mass of the solution  is determined by
$c_{\pm}, n$. This can be explicitly verified using in the previous equation 
the asymptotic expansion for the scalar field (\ref{ae}) and that for 
$g_{tt}$,
\be\lb{me}
e^{2f}= \frac{r^{2}}{L^{2}}- M +{\cal O}(\frac{1}{r}),
\ee
one immediately finds that $M$   is determined, algebraically, by $c_{\pm}$
and $n$. If only  a single fall-off mode for the scalar (only one 
independent scalar charge $c$) is
present in Eq. 
(\ref {ae}), the  black hole mass $M$  is completely determined by 
the scalar charge $c$ and by the winding number $n$.
\section{Linear form of the field equations}
In order to find exact solutions of the system (\ref{4.14}) one has to 
choose a form of the potential $V(\vf)$ for the scalar field.
Alternatively, one can choose a form for the scalar field $\vf(r)$
and solve (\ref{4.14}) in terms of $h(r),f(r), V(r)$.
This method has two main advantages. It allows to fix from the 
beginning the asymptotic behavior of the field $\vf$.  The field 
equations
(\ref{4.14}) can be  rewritten in the form of a system of 
first order linear differential 
equations.
This can be achieved introducing in Eqs. (\ref{4.14}) 
the new variables $S,Z$
\be\lb{nv}
S=\alpha e^{-2 h},\quad Z= \alpha( V+\frac{1}{2}\frac{n^{2}}{r^{2}}
 \vf^{2}-Y_{0}¥),
\ee
where  $\alpha(r)=e^{\int dr r {\dot \vf}^{2}}$ and $Y_{0}$ is a solution 
of the equation $\dot{Y_{0}} +r {\dot \vf}^{2}Y_{0}+
\frac{n^{2}}{r^{3}}
 \vf^{2}=0$.
 After eliminating $f$ and using Eqs. (\ref{nv}),
 the second and third equations in  (\ref{4.14}) 
 can be written in the form
 \be\lb{le}
 \dot{S} -p(r) S+r Z+q(r)=0,\quad \dot{Z} -g(r) S=0,
 \ee
where $p(r)=\frac{1}{2} r({\dot \vf})^{2},\, q(r)= r \alpha Y_{0},\,
g(r)= \frac{1}{2r^{2}}(d/dr)(r^{2}{\dot \vf}^{2}).$
This is a first order system of linear differential equations, which 
allows to determine the potential $V$ and the metric function $h$ once 
 $\vf(r)$ is given. The metric function $f$ then follows just by 
 integrating the first equation  in (\ref{4.14}).
 
Although this method  for solving Eq. (\ref{4.14}) 
can be used for a generic $\vf(r)$, in this paper 
we will mainly  apply it 
to the case of  a 
scalar field that saturates the BF 
bound in 3D, this implies $\lambda_{+}=\lambda_{-}=1,\, 
m^{2}=-1/L^{2}=m_{BF}^{2}¥$, i.e \footnote{ We do not consider the logarithmic 
branch (see footnote 1)}
\be\lb{ccc}
\vf=\frac{c}{r},\quad  V^{\prime \prime }(0)=\frac{1}{2}V(0).
\ee
In the next sections we will derive in closed form exact  
 black hole 
and vortex solutions  corresponding a scalar field given by Eq. 
(\ref{ccc}).
We will also briefly   consider  solutions of
Eqs. (\ref{4.14}) in
the case of  scalar field with 
$m^{2}¥=0$, i.e for $\vf=c /r^{2}$.

\section{Black hole solutions with real scalar hairs}
For a  real scalar field, i.e. for $n=0$, and $\vf$ given by Eq. (\ref{ccc})
the method described in the previous section allows us to find the 
following  solution of 
the system (\ref{4.14}), 
\bea\lb{neqz}
V(\vf )&=&-2\lambda_{1}e^{\frac{%
\vf ^{2}}{2}}+\frac{\lambda_{1}}{2}\vf ^{2}e^{\frac{\vf ^{2}}{2}}-2\lambda_{2}e^{%
\frac{\vf ^{2}}{4}},\nonumber\\
e^{-2h}&=&r^{2}e^{\frac{c^{2}}{4r^{2}}}\left(\lambda_{1} e^{\frac{c^{2}}{4r^{2}}}
+ \lambda_{2}\right),\nonumber\\
e^{2f}&=&e^{-\frac{c^{2}}{2r^{2}}}e^{-2h}.
\eea
where $\lambda_{1,2}$  are parameters entering in the potential. They 
define the cosmological constant of the AdS spacetime. In fact using 
Eq. (\ref{cc}) one finds
\be\lb{ll}
L^{-2}=\lambda_{1}+\lambda_{2}.
\ee
Thus, AdS solution exist only for $\lambda_{1}>-\lambda_{2}$. 
For $\lambda_{1}>0$  the potential   has the $W$ form typical of 
Higgs potentials. The maximum  of the potential,   
corresponds to an AdS$_{3}$ vacuum  with $\vf=0$ and 
$V(0)=-2/L^{2}¥$. It preserves   the global $U(1)$ symmetry  
of the action (\ref{4.1}). Conversely, the minimum 
of the potential corresponds to an other AdS$_{3}$ vacuum with 
$\vf=\vf_{m}$ and  a different AdS length $l$, $-2 l^{-2}=V( 
\phi_{m})$. This vacuum  
breaks
spontaneously the $U(1)$ symmetry of the action (\ref{4.1}).
The shape of the potential for $\lambda_{1}¥ >0$ is shown in figure 1 
(left).

For $\lambda_{1}<0$ the potential has only the 
symmetry preserving maximum at $\vf=0$ and no symmetry breaking minima.
The shape of the potential for 
$\lambda_{1} <0,\lambda_{2}>0,\lambda_{2}>|\lambda_{1}|
$ is shown in figure 1 (right).

Let us now consider the metric part of the solution (\ref{neqz}).
The solution describes a black hole only for 
$\lambda_{1}<0,\lambda_{2}>0 $, which in view of Eq. (\ref{ll}) 
requires $\lambda_{2}>|\lambda_{1}|$.
As expected, the black hole mass $M$ is determined in terms of the 
scalar charge $c$ by the asymptotic expansion (\ref{me}) of $e^{2f}$, 

\be\lb{M}
M=\frac{\lambda _{2}c^{2}}{4}.
\ee
The event horizon is located at 
\be\lb{bh}
r_{h}=\gamma\sqrt M, \quad 
\gamma=\left[\lambda_{2}
\ln(-\frac{\lambda_{2}}{\lambda_{1}})\right]^{-\frac{1}{2}}¥.
\ee
$r=0$ is a curvature singularity with the scalar curvature 
behaving as \,\,$R\sim \exp({2M/(r^{2}\lambda_2)}) /r^{4}$.

For $\lambda_{1}>0$ the solution (\ref{neqz}) describes a naked singularity.
Because black hole solutions exist only in the range of parameters 
for which the potential $V$ has no minima, it follows that there 
is no $n=0$ black hole solution interpolating between 
the $U(1)$-symmetry preserving vacuum at $\vf=0$ and the symmetry 
breaking vacuum at $\vf=\vf_{m}$.
The symmetry-preserving and the symmetry-breaking vacua seem 
therefore gravitationally disconnected.
Black hole solutions with scalar hairs exist only for the potential 
shown in right hand side of figure 1. 

\begin{figure}
    \includegraphics{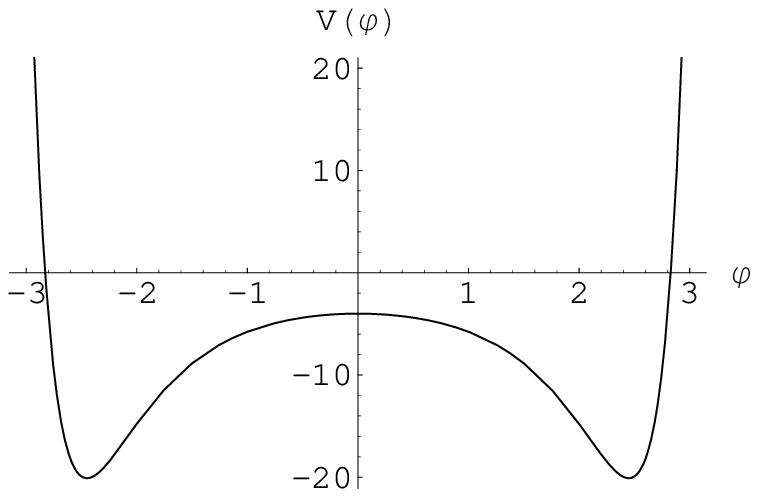} \includegraphics{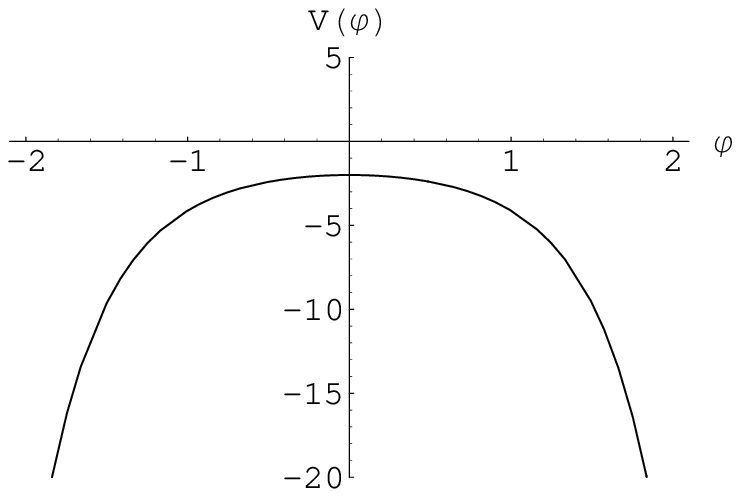}\ 
  \caption{
  {\sl Left.} The form of the potential $V(\vf)$ for the case 
    $\lambda_{1}, \lambda_{2}>0$. The picture  shows the 
    potential for  $\lambda_{1}=\lambda_{2}=1$.
    {\sl Right.} The form of the potential $V(\vf)$ for the case 
    $\lambda_{1}<0, \lambda_{2}>0,\lambda_{2}>|\lambda_{1}|$. 
    The picture  shows the 
    potential for  $\lambda_{1}=-1, \lambda_{2}=2$.}
  \label{fig:dna}
\end{figure}

It is also interesting to have a short  look at the thermodynamics of the black 
hole described by the solution (\ref{neqz}).
Temperature and entropy are given by
\be\lb{te}
T=\frac{1}{2\gamma\pi} \sqrt{M},\quad S=4 \pi \gamma \sqrt M,
\ee
where $\gamma$ has been defined in Eq. (\ref{bh}).
One can easily show that $M,T, S$ satisfy the first principle, 
$dM=TdS$. The thermodynamical behavior of our black hole follows 
closely that of the BTZ black hole, which is given by the same 
equations (\ref{te}) with $\gamma=L$.
If we compute the free energy of the BTZ black hole  and that  of the 
scalar-hair  black hole 
(\ref{neqz}) one finds respectively:
\be\lb{free_energy}
F_{BTZ}= -4\pi L^{2}T^{2},\quad F_{SH}= -4\pi \gamma^{2}T^{2}.
\ee

Because in the range of existence of the black hole solutions 
($\lambda_{2}>|\lambda_{1}|$), we have always $\gamma<L$, it follows
 that $F_{SH}>F_{BTZ}$. This is consistent with the stability the BTZ 
 black hole in the presence of a real scalar field with mass 
 satisfying the BF bound and with the interpretation of 
 the scalar hair solution (\ref{neqz}) as an excitation of AdS$_{3}$.

\section{Vortex  solutions}
Before considering vortex-like, black hole solutions, let us first 
consider the vacuum, $n=0$, solutions of a model with   
a potential for the 
scalar  of the form:
\be\lb{pot}
V(\vf )=-4\Lambda^{2}-\Lambda^{2}\vf^{2}-\frac{\Lambda^{2}}{2}\vf^{4}¥ 
-2\lambda_{1}e^{\frac{%
\vf ^{2}}{2}}+\frac{\lambda_{1}}{2}\vf ^{2}e^{\frac{\vf ^{2}}{2}}-2\lambda_{2}e^{
\frac{\vf ^{2}}{4}},
\ee
where $\Lambda,\lambda_{1}>0,\lambda_{2}<0$ are parameters.
Also in this case the maximum  of the potential,   
corresponds to a AdS$_{3}$, $U(1)$-symmetry preserving vacuum,  with $\vf=0$ and
\be\lb{f2}
V(0)=-\frac{2}{L_{1}^{2}}=-4\Lambda^{2}-2\lambda_{1}-2\lambda_{2},
\ee
where as usual $L_{1}$ is the AdS length. Existence of this AdS$_{3}$ vacuum obviously requires 
$2\Lambda^{2}+\lambda_{1}+\lambda_{2}>0$.

The mass of the scalar excitation near $\vf=0$ saturates the BF bound, 
in fact we have $m^{2}=V''(0)=(1/2) V(0)= -1/L_{1}^{2}=m_{BF}^{2}.$
We have also a minimum of the potential describing a $U(1)$-symmetry 
breaking AdS$_{3}$ vacuum at  $\vf= \hat \vf_{m}$ with AdS length 
$l_{1}$ given by $-2 l_{1}^{-2}=V( \hat \vf_{m}).$

Using the method described in Sect. 4, a $n\neq 0$ vortex-like, black hole  solution of the field equations 
(\ref{4.14}) can be  obtained for a profile of the complex scalar 
field given by
\be\lb{csf}
\phi= \frac{c}{r} e^{i n\theta},
\ee
when the parameter 
$\Lambda$ of the potential is fixed in terms of $c$ and $n$: 
$\Lambda=n/c$ and $ \Lambda^{2}¥\leq\frac{\lambda_{2}^{2}}{8 \lambda_{1}}$.

The potential (\ref{pot}) with $\Lambda=n/c$ has the physical meaning 
of the effective potential seen by a vortex with scalar charge $c$ 
and winding number $n$.
The metric part of the black hole solution turns out to be,
\be\lb{vbh}
e^{-2h}=\lambda_{1}r^{2}e^{\frac{c^{2}}{2r^{2}}}+\lambda_{2}r^{2}e^{\frac{c^{2}}{4r^{2}}%
}+\frac{2n^{2}}{c^{2}}r^{2},\quad e^{2f}=e^{-\frac{c^{2}}{2r^{2}}}e^{-2h}.
\ee
The black hole mass is given in terms of the scalar charge $c$ and 
the winding number $n$:
\be\lb{bhm}
 M=\frac{\lambda _{2}c^{2}}{4}+n^{2}.
\ee
The black hole solution (\ref{vbh}) has an outer and inner horizon 
located respectively at

\be\lb{hor}
r_{\pm} =\frac{c}{2}\left(\ln \left\{ 
\frac{1}{2\lambda_{1}}\left[ -\lambda_{2}\mp
\sqrt{\lambda_{2}^{2}-\frac{8\lambda_{1}n^{2}}{c^{2}}}\right] 
\right\} \right)^{-\frac{1}{2}}.¥\ee
The black hole solution exists for 
\be\lb{bound}
\frac{M}{n^{2}}\leq 1-2\frac{\lambda_{1}}{|\lambda_{2}|},\,\qquad 
|\lambda_{2}|>2 \lambda_{1}.
\ee
The black hole becomes extremal (single event horizon) when the 
bound (\ref{bound}) is saturated.
In general the radial coordinate $r_{m}$ corresponding to
the minima of the potential (\ref{pot}) can 
lie either outside or inside the event horizon. The requirement  $r_{m}\ge 
r_{h}$ implies $M/n^{2}\ge 1-(9/4) (\lambda_{1}/|\lambda_{2}|)$.

Thus the vortex black hole solution (\ref{vbh}) interpolates between a 
 symmetry-preserving AdS$_{3}$ vacuum at $\vf=0$  and a symmetry 
 breaking AdS$_{3}$ vacuum at $\vf=\vf_m$ for
 \be\lb{sss}
 1-\frac{9\lambda_{1}}{4|\lambda_{2}|}\leq\frac{M}{n^{2}}
 \leq1-2 \frac{\lambda_{1}}{|\lambda_{2}|}.
 \ee
 Notice that differently from the usual case the vortex 
 connects a maximum of the potential in the $r\to\infty$ 
 asymptotic region with a minimum of the potential in the interior 
 region.  
 This is related with the fact that the energy of the vortex although 
 finite is always negative.
 The energy $E$ of 
 the vortex is given by
 \be\lb{energy}
 E=\int_{r_{+}}^{\infty}r dr d\theta(T_{t}^{t}) - E_{\text{vacuum}},
 \ee
 where $T_{t}^{t}$ is the stress energy tensor of the complex scalar 
 field $\phi$ and we have subtracted the contribution of the 
 $\phi=0$  vacuum.
 We find after  a little algebra,
 \be\lb{st}
 E=\frac{c^{2}}{4} (\lambda_{2}+2 \lambda_{1})= (M-n^{2})(1-2
  \frac{\lambda_{1}}{|\lambda_{2}|}).
 \ee
 The energy of the vortex measured with respect to the vacuum
 is finite and always  negative.
\subsection{Stability of the  solutions}
Let us first consider the solutions with  a real scalar field.
In this case the potential $V(\vf)$, given in Eq. (\ref{neqz}) with
$\lambda_1<0$, admits as solution both the BTZ black hole and a  
dressed black hole
with a real scalar field
$\vf=c/r$. The dressed black hole always has a free energy which is larger than that of the BTZ
black
hole. Correspondingly, the dressed black hole is always unstable and it will decay to the stable BTZ
black hole, losing its real scalar hair. This is confirmed by a linear stability analysis. In the BTZ
background scalar perturbations, $\delta\phi$, decouple from the metric. 
The equation of motion for
the scalar perturbation
is  given by the Klein-Gordon equation for a scalar field propagating in AdS with a mass
$m^2=-1/L^2$ which saturates the BF bound of $AdS_3$. Thus the BTZ solution is (at least linearly)
stable. 

The story does not change so much if we consider solutions with a 
complex scalar field. In this case  the potential $V(\vf)$, supporting
vortex solutions, $\phi=(c/r) e^{in\theta}$ in given in Eq. (\ref{pot}).
We can consider scalar perturbations around the BTZ
black hole, which is again solution of the equations of motions.
Because  the potential only depends on $\vf=|\phi|$, at the linear 
level the equation for scalar
perturbations near  extrema of the potential always decouple in a 
holomorphic and antiholomorphic part:
\ba
&&\frac{1}{\sqrt{-g}}\partial _{\mu }(\sqrt{-g}\partial ^{\mu }\delta\phi)+
\frac{1}{L^2}\delta\phi
 =0,\nonumber\,,\\
&&\frac{1}{\sqrt{-g}}\partial _{\mu }(\sqrt{-g}\partial ^{\mu
}\delta\bar{\phi)}+\frac{1}{L^2}\delta\bar{\phi}\,,
 =0,\nonumber
\ea
i.e. we have two independent real scalar  perturbations,
propagating in the BTZ background  with the  same mass saturating the 
BF bound,
$m^2=-1/L^2=m^2_{BF}$. This  again ensures  linear stability of the BTZ black hole.

\subsection{ Other Vortex solutions}
Let us now briefly  consider a scalar field with $m^{2}=V''(0)=0$. From 
equation (\ref{ae}) it follows that $\vf$ must  have the 
form
$\vf =\frac{c}{r^{2}}$. We will therefore look for a complex scalar field   
 of 
the form: 
\begin{equation}
\phi =\frac{c}{r^{2}}e^{i n\theta}.
\end{equation}
In this case the method described in Sect. 4 gives  the following 
solution for the potential $V(\vf)$,
\bea\lb{pot1}
V(\vf ) &=&-2\lambda_{1}e^{\vf ^{2}}-\frac{n^{2}}{2c}\vf +
2\sqrt{\frac{2}{\pi }}%
\lambda_{2}\vf e^{\frac{\vf ^{2}}{2}}+2\lambda_{1}\vf ^{2}e^{\vf ^{2}}+ 
\notag \\
&&\frac{n^{2}\sqrt{\pi }}{4c}\vf ^{2}e^{\vf ^{2}}\left[ 1-\erf\left(
\vf \right) \right] -\frac{n^{2}\sqrt{\pi }}{4c}e^{\vf ^{2}}\left[ 
1-\erf\left( \vf \right) \right]   \notag \\
&&-2\lambda_{2}e^{\vf ^{2}}\erf\left( \frac{\vf }{\sqrt{2}}\right)
+2\lambda_{2}\vf ^{2}e^{\vf ^{2}}\erf\left( \frac{\vf }{\sqrt{2}}\right) -%
\frac{n^{2}}{2c}\vf ^{3},
\end{eqnarray}%
where $\lambda_{1},\lambda_{2}$ are constants.

The metric part of the solution is given by
\begin{equation}\lb{mp}
 e^{2f}=r^{2}\left\{\lambda_{1}+\lambda_{2}\erf\left( \frac{c}
 {\sqrt{2}r^{2}}\right) +%
\frac{n^{2}\sqrt{\pi }}{8c}\left[ 1-\erf\left( \frac{c}{r^{2}}%
\right) \right]\right\},\, 
e^{-2h}=e^{\frac{c^{2}}{r^{4}}} e^{2f},
\end{equation}%
whereas for 
the AdS length we have,
\begin{equation}
V(0)=-\frac{2}{L^{2}} =-2\lambda_{1}-\frac{n^{2}\sqrt{\pi }}{4c}.
\end{equation}%
Hence, the solutions are asymptotically  AdS for 
$\lambda_{1}> -\frac{n^{2}\sqrt{\pi }}{8c}$.

The mass $M$ of the solution can be easily computed using the asymptotic 
expansion of $\exp(2 f)$. One has,
\be\lb{ma}
M= - \sqrt{\frac{2}{\pi}}\lambda_{2} c+\frac{n^{2}}{4}.
\ee
The question about the presence of event horizon in the generic, 
$n\neq 0$, solution given by Eq.  (\ref{mp}) is rather involved. We will not address 
this problem here, but we will just consider the $n=0$ solutions.
The solutions are asymptotically AdS for $\lambda_{1}>0$.
They describe a black hole for 
$\lambda_{2}<0,\,|\lambda_{2}|\geq \lambda_{1}$.
The position of event horizon $r_{h}$ is  the solution of the equation
$$\erf(\frac{c}{\sqrt{2}r_{h}¥^{2}})= 
\frac{\lambda_{1}}{|\lambda_{2}|}.$$
\section{Conclusions}
In this paper we have derived and discussed exact, spherically 
symmetric, solutions  of 3D 
AdS gravity coupled with a complex scalar field $\phi$.
Our  method  allows to determine the potential $V(|\phi|)$  
once the form of $|\phi(r)|$ is fixed.
A scalar field profile corresponding to the saturation of the BF 
bound requires a Higgs-like potential $V$, which allows both for BTZ 
black holes solutions - corresponding to a constant scalar 
field- and 
for black hole solutions with  scalar hairs. 

A generic  
feature of 3D black holes with scalar hairs is that
the black hole mass is determined algebraically by 
the 
scalar charges and by the winding number of the vortex.
The main difference between
3D black hole solutions with real scalar hairs and the BTZ black hole 
is the presence of a curvature singularity at  $r=0$, where the 
scalar curvature behaves as $R\sim e^{2M/(r^{2}\lambda_2)}/r^{4}$.
On the other hand, the thermodynamical behavior of these dressed 
black hole solutions is very similar to that of the BTZ black hole. 
They share the same mass/temperature relation $M\sim T^{2}$  and we 
were able to show that the BTZ black hole remains always stable, 
being the free energy of the dressed solution 
always bigger than that of the BTZ black hole. 

Conversely,  the  black hole-vortex 
solutions we have derived in this paper share many feature with both 
rotating \cite{Hotta:2008xt} and electrically charged \cite{Cadoni:2008mw} 
3D black hole solutions.
They are characterized by the presence of an inner and outer horizon
and the vortex interpolates between two AdS$_{3}$ geometries with 
different AdS 
lengths. For some choice of the parameters, these 
AdS$_{3}$ geometries correspond respectively  to the 
$U(1)$-symmetry-preserving 
maximum  and to the symmetry-breaking minimum of the Higgs potential.
In the AdS/CFT language this means that the vortex interpolates 
between two 2D CFTs with different central charges. 
It is interesting to notice that these interpolating  features of the 
vortex solution are more similar to those of the rotating solution of 
Ref. \cite{Hotta:2008xt} (both interpolated geometries are AdS$_{3}$) 
then to those of the electrically charged solutions of 
Ref. \cite{Cadoni:2008mw} (the solution interpolates between AdS$_{3}$ and 
AdS$_{2}\times S^{1}$).

Also the non trivial solution for the scalar field could have an 
interesting holographic interpretation. The scalar charge $c$ has to 
be thought of as a nonvanishing  VEV for some boundary operator. 
This could be a signal of a  phase transition in  
the dual 2D boundary field theory. However, this is a rather 
involved  issue because  of the well-known results about   phase transitions 
in  statistical systems with only one spacelike dimension \cite{ruelle}.

\end{document}